\documentclass[preprint,preprintnumbers,amsmath,amssymb,nofootinbib,superscriptaddress]{revtex4}

\usepackage{graphicx}
\usepackage{dcolumn}
\usepackage{bm}

\begin{document}


\title{AdS/NRCFT for the (super) Calogero model}

\author{Wen-Yu Wen}
\email{steve.wen@gmail.com}
\affiliation{Department of Physics and Center for Theoretical Sciences\\ 
National Taiwan University, Taipei 106, Taiwan}


\begin{abstract}
We propose a correspondence between the non-relativistic quantum Calogero model in $d$ spatial dimensions and classical gravity theory in a deformed $AdS_{d+3}$ background for single particle and $AdS_{Nd+3}$ for $N$-particles.  Later we extend our construction to the superspace for a super Calogero model.   
\end{abstract}


\maketitle

\section{Introduction}
Recently the AdS/CFT corrrespondence, originally proposed for supersymmetric conformal field theory\cite{Maldacena:1997re,Gubser:1998bc,Witten:1998qj}, has been established for the non-relativistic conformal field theory\cite{Son:2008ye,Balasubramanian:2008dm}, denoting as the AdS/NRCFT correspondence.  This may cast new insights into earlier works on realization of Schr\"{o}dinger group\cite{Hagen:1972pd,de Alfaro:1976je} in terms of isometry\cite{Burdet:1977qw,Duval:1984cj,Henkel:2003pu}.  The new feature in this correspodence is the explicit realization of conformality by some deformed AdS geometry in a holographic way.  In addition to the AdS vacuum, some kind of pressureless matter seems necessary for generating the desired metric which respects the Schr\"{o}dinger group.  Nevertheless, it was pointed out by \cite{Goldberger:2008vg,Barbon:2008bg} that without this exotic matter, the correspondence is still well established up to the discrepancy in the definition of mass appearing in the Schr\"{o}dinger equation.  Moreover, the AdS/NRCFT correspondence can be easily applied to the system with a harmonic trap potential.  This application is gurranteed to be successful mainly thanks to the unbroken $SL(2,R)$ symmetry hidden inside the whole $SO(d+4,2)$ conformal group, such that one is free to redefine its Hamiltonian to include a harmonic potential term by deforming the metric component $g_{tt}$ accordingly.  In this paper, we would like to make a full use of this $SL(2,R)$ symmetry and propose a general holographic metric corresponding to a most general Hamiltonian found in the quantum Calogero model.  Our proposed metric includes additional deformation of component $g_{t\vec{x}}$ and we have studied its role in the AdS/NRCFT correspondence.  Later we explore the possibility to extend our construction to the superspace corresponding to a super Calogero model.  This paper is organized as follows: in the section II we briefly review the quantum Calogero model.  In the section III we propose a deformed $AdS_{d+3}$ metric corresponding to single particle and $AdS_{Nd+3}$ for $N$ particle states, with identical or different masses.  We also extend our construction to the superspace in order to describe a super Calogero model.  At last, we have some discussion and remarks in the section IV.

\section{The Calogero model}
One-dimensional $N$-particle Calogero model was first studied by \cite{Calogero:1969xj,Perelomov:1971ki} as one of those rare  exactly solvable many-body quantum mechanical systems.  In that model, a particle interacts with each other via a combination of an attractive harmonic force and a repulsive inverse square potential.  Here we are interested in a general $N-$particle Calogero model in arbitrary $d+1$ dimenions described by a Hamiltonian in the unit $\hbar=1$, following \cite{Meljanac:2005hm}
\begin{equation}\label{hamiltonian_calogero}
{\cal H}=-\frac{1}{2}\sum{\frac{1}{m_i}\nabla_i^2}+V(\vec{x_1},\dots,\vec{x_{N}})+\frac{\omega^2}{2}\sum{m_i\vec{x_i}^2}+\frac{c}{4}(\sum{\vec{x_i}\nabla_i+\nabla_i\vec{x_i}}),
\end{equation}
where we assume the potential is invariant under translation and a homogeneous function of order $-2$ such that
\begin{equation}
[\sum{\vec{x_i}\nabla_i,V}]=-2V.
\end{equation}
It is instructive to arrange the Hamiltonian in the following way,
\begin{eqnarray}\label{algebra_sl2}
&&{\cal H}=-T_-+\omega^2T_++cT_0,\nonumber\\
&&T_+ = \frac{1}{2}\sum{m_i\vec{x_i}^2},\nonumber\\
&&T_- = \frac{1}{2}\sum{\frac{1}{m_i}\nabla_i^2}-V(\vec{x_1},\dots,\vec{x_{N}}),\nonumber\\
&&T_0 = \frac{1}{4}(\sum{\vec{x_i}\nabla_i+\nabla_i\vec{x_i}}),
\end{eqnarray}
where the $SL(2,R)$ conformal symmetry manifests via 
\begin{equation}
[T_-,T_+]=2T_0,\qquad [T_0,T_{\pm}]=\pm T_{\pm}.
\end{equation}
The above Hamiltonian can be viewed as a deformation of $N$ harmonic oscillators in $d$ spatial dimensions.  In generic, the common frequency $\omega$ could be different for different species of particles.  Solutions to the above Hamiltonian are irrelevant to our concern in this paper so we will skip those details which can be found in \cite{Meljanac:2005hm}.  We would like to emphasize that these generators $\{ T_0,T_{\pm}\}$ can be casted into sum of the center-of-mass (CM) and relative parts, while the same separation is also true for the energy.  Later we will find out that exactly the same separation appears while we establish the correspondence.

\section{Holographic realization of Calogero model}
In this section, we would like to construct a gravitional model holographically dual to the non-relativistic Calogero model mentioned in the previous section.  We will begin with one-particle state and end with $N$-particle of different species.  Later we will try to extend this construction to the superspace and propose a candidate which corresponds to a non-relativistic super Calogero model.

\subsection{Single particle state}
We start with one particle with mass $m$ for simplicity.  In order to embed a  non-relatvistic model into a relativistic gravity, one option is to work with additional two dimensions spanned by a holographic direction $r$ and a compactified null direction $\xi$\cite{Son:2008ye}. 
We propose a general ansatz for the metric:
\begin{equation}\label{metrics_one}
ds^2=-2 \frac{r^4}{R^4}A(r,\vec{x})dt^2+\frac{r^2}{R^2}(-2dt d\xi + d\vec{x}^2)+\frac{r^3}{R^3}(-2dt d\vec{x}\cdot \vec{B}(r,\vec{x}))+\frac{R^2}{r^2}dr^2,
\end{equation}
where $R$ used to be the curvature radius of pure $AdS_{d+3}$ space but now is simply taken as a dimensional constant to make $g_{\mu\nu}$ dimensionless.  Notice that we have asked our metric and functions $A,\vec{B}$ to respect symmetry of the Schr\"{o}dinger group, that is
\begin{equation}
(t,\vec{x},\xi,r)\to (\lambda^2 t, \lambda \vec{x}, \xi,\lambda^{-1} r).
\end{equation}
For our purpose here, we will restrict functions $A,\vec{B}$ to take the form:
\begin{equation}\label{ansatz_one}
A(r,\vec{x})=\gamma^2 (\frac{R}{r})^4 + a_{-2} (\frac{R}{r})^2\frac{1}{|\vec{x}|^2} + a_0 + a_2 (\frac{R}{r})^2 |\vec{x}|^2,\qquad \vec{B}(r,\vec{x})=b_1 (\frac{R}{r}) \vec{x},
\end{equation}
where dimensionality of all coefficients is chosen such that whole functions are invariant under $\lambda$ rescaling.
Now we turn to the study of the correspondence. In particular, we consider an operator ${\cal O}(t,\vec{x})$ dual to a massive scalar field $\phi(r,\xi,t,\vec{x})$ with mass $m_0$.  The minimal coupled action reads,
\begin{equation}
S=-\int{d^{d+3}x}\sqrt{-g}(g^{\mu\nu}\partial_\mu\phi^*\partial_\nu\Phi+m_0^2\Phi^*\Phi)
\end{equation}
We will further impose the condition $-i\partial_\xi =M$ for a compactified $\xi$ direction of size $1/M$, and request the scalar field to take the stationary form:
\begin{equation}
\Phi(r,\xi,r,\vec{x})=e^{i\omega t+iM\xi} \Lambda(r)\Psi(\vec{x}).
\end{equation}
Then the equation of motion for $\Phi$ can be casted into two parts: the part which only depends on $\vec{x}$ reads
\begin{equation}
-\frac{1}{2M}\nabla^2\Psi + M\frac{a_{-2}}{|\vec{x}|^2}\Psi + \frac{M}{2}\omega^2 |\vec{x}|^2 \Psi + i b_1 \vec{x}\cdot\nabla \Psi = \epsilon_x \Psi,\qquad \omega^2\equiv 2a_2+b_1^2.
\end{equation}
$\Psi(\vec{x})$ is ready to be solved once one recognizes that this is the very one-particle Calogero model mentioned in the previous secion, where $c=i 2b_1$ and $\epsilon_x$ is the Hamltonian up to a constant $d/4$.

The part which only depends on $r$ reads,
\begin{eqnarray}
&&-\frac{1}{2M}[\nabla_z^2\Lambda - \frac{d+1}{z}\nabla_z\Lambda] + [\frac{m^2}{2Mz^2} +\frac{M\gamma^2}{2}z^2 - \epsilon_z]\Lambda=0,\nonumber\\ 
&&\epsilon_z\equiv E-\frac{i}{2}db_1-\epsilon_x,\qquad m^2 \equiv m_0^2+2a_0M^2,
\end{eqnarray}
where we have made a coordinate transformation $z=R^2/r$.  A special case $\gamma=b_1=0$ was studied in the \cite{Son:2008ye}.  In generic, $\Lambda(r)$ takes a linear combination of confluent hypergeometric and generalized Lagruerre function:  
\begin{eqnarray}
&&\Lambda(z)=2^{\frac{1+\nu}{2}}e^{\frac{1}{2}M\gamma z^2}z^{\frac{d}{2}+1+\nu}\{c_- U(n,1+\nu,-M \gamma z^2)+c_+ L_{-n}^\nu(-M\gamma z^2)\},\nonumber\\
&&n \equiv \frac{1+\nu}{2}+\frac{\epsilon_z}{4M\gamma},\qquad \nu \equiv \sqrt{m^2+(\frac{d+2}{2})^2}.
\end{eqnarray}
Note that for asymptotic boundary $z\to 0$, $\Lambda(z)\sim c_- z^{\Delta_-}+c_+ z^{\Delta_+}$, where $\Delta_{\pm}=\frac{d+2}{2}\pm \nu$.  As in usual AdS/CFT correspondence, one considers an operator $\cal O$ dual to this massive scalar field $\Phi$.  For the choice of $\nu\le 1$, only the first term is renormalizable threrfore the second term acts like a source.  The correlation function then reads
\begin{equation}
<{\cal O}{\cal O}>\sim |\epsilon_z|^{2\nu},
\end{equation}
where the scaling dimension of $\cal O$ is $\frac{d+2}{2}+\nu$.  However, for $1 > \nu >0$, both solutions are renormalizable and one is free to choose either one as the source and the other as the condensate.

If we are careful enough to tune the parameter $\gamma=\omega$, one is tempted to identify $z$ as the internal hyperradius and $\epsilon_z$($\epsilon_x$) as the internal energy (the CM energy).  It has been known that the total Hamiltonian $E$ in this conformal system can be nicely separated into the CM part $\epsilon_x$ and internal part $\epsilon_z$\cite{Werner:2006}.  We have, in addition, another contribution $-cd/4$ due to dilation operator $T_0$.  In summary, our proposed metric (\ref{metrics_one}) and (\ref{ansatz_one}) successfully reproduce equations for both CM and internal motions for the one-particle Calogero model.  Interpretating $z$ as the hyperradius results to $z^2=|\vec{x}|^2$.  We immdiately learn that low(high) energy scale $z$ corresponds to small(large) fluctuation $\delta\vec{x}$ is the very spirit of holography.

\subsection{$N$-particle state}
In this section, we consider a Calogero model of $N$ identical particles.  The proposed metric in $Nd+3$ dimensions reads,
\begin{eqnarray}\label{metrics_two}
ds^2=&&-2 \frac{r^4}{R^4}A(r,\vec{x_i})dt^2+\frac{r^2}{R^2}(-2dt d\xi + \sum_i^N{d\vec{x_i}^2}) \nonumber\\
&&+\frac{r^3}{R^3}(-2\sum_i^N dt d\vec{x_i}\cdot \vec{B_i}(r,\vec{x_i}))+\frac{R^2}{r^2}dr^2, \nonumber\\
\end{eqnarray}
and 
\begin{eqnarray}
&&A(r,\vec{x_i})=\gamma^2 (\frac{R}{r})^4 + (\frac{R}{r})^2\sum_{i\neq j}{\frac{a_{-2}}{|\vec{x_i}-\vec{x_j}|^2}} + a_0 + a_2 (\frac{R}{r})^2 (\sum_i^N{|\vec{x_i}|^2}),\nonumber\\ 
&&\vec{B_i}(r,\vec{x_i})=b_1 (\frac{R}{r}) \vec{x_i}.
\end{eqnarray}
Here $\vec{x_i}$ has the interpretation of position of the $i$th particle and the hyperradius  $z=R^2/r=\sqrt{|\sum_i^N{\vec{x_i}|^2}}/N$.  We again impose the condition $-i\partial_\xi = NM$ for compactified $\xi$ direction.  Following a similar procedure for single particle, one can easily reproduce the Calogero model of $N$ identical particles with energy reltation $NE=N\epsilon_x+N\epsilon_z+\frac{i}{2}Ndb_1$.

\subsection{particles with different masses}
The $N$-particle state can be further generalized to the case of particles with different masses.  This is done by observing that all $m_i$'s can be absorbed by redefinition of $\vec{x_i}$\footnote{See also the discussion in the Appendix A in the \cite{Werner:2006}.}.  Indeed, this can be realized by imposing $-i\partial_\xi=1$ (compactified circle of unit mass) and 
\begin{eqnarray}
&&A(r,\vec{x_i})=\gamma^2 (\frac{R}{r})^4 +  (\frac{R}{r})^2\sum_{i\neq j}{\frac{a_{-2}^{ij}}{|\vec{x_i}-\vec{x_j}|^2}} + a_0 \sum_i^N{m_i^2} + a_2 (\frac{R}{r})^2 (\sum_i^N{m_i|\vec{x_i}|^2}),\nonumber\\ 
&&\vec{B_i}(r,\vec{x_i})=b_i (\frac{R}{r}) m_i\vec{x_i},
\end{eqnarray}
where we have allowed more paramters $a_{-2}^{ij}$ and $b_i$ for particles of different masses.  We then have to redefine 
\begin{equation}
m^2=m_0^2+\sum_i^N{m_i^2},\qquad \omega_i^2=2a_2+b_i^2,
\end{equation}
for the effective mass of bulk scalar and angular frequency for the $i$th particle.

\section{super Calogero model}
In this section, we would like to formulate a holographic model dual to a super one-particle Calogero model in the superspace $R^{d|2n}$ for $d$ commuting coordinates $x^i$ and $2n$ anti-commuting ones $\grave{x}^j$.   Following \cite{De Bie:2007uc}, the spatial coordinate $\vec{x}$ is extended in the following way,
\begin{equation}
{\tilde x}=\sum_{i=1}^d{x^i e_i}+\sum_{j=1}^{2n}\grave{x}^j\grave{e_j}, 
\end{equation}
where $e_i$ and $\grave{e}_j$ are orthogonal and sympletic basis satisfying
\begin{eqnarray}
&&\{ e_j,e_k \}=-2\delta_{jk},\qquad \{ e_j,\grave{e}_k \}=0,\nonumber\\
&&[\grave{e}_{2j},\grave{e}_{2k}]=0,\qquad [\grave{e}_{2j-1},\grave{e}_{2k-1}]=0,\qquad [\grave{e}_{2j-1},\grave{e}_{2k}]=\delta_{jk}.
\end{eqnarray}
such that the {\it distance} measured in the superspace is
\begin{equation}
d{\tilde x}^2=-\sum_{i=1}^d{(dx^j)^2}+\sum_{j=1}^n{d\grave{x}^{2j-1}d\grave{x}^{2j}}
\end{equation}
Now we refine the $SL(2,R)$ generators as:
\begin{eqnarray}
&&\tilde{T}_+= \frac{1}{2}m {\tilde x}^2 = \frac{1}{2}m(\sum_{j=1}^n{\grave{x}^{2j-1}\grave{x}^{2j}}-\sum_{i=1}^d{(x^i)^2}),\nonumber\\
&&\tilde{T}_-= \frac{1}{2m}\nabla_{\tilde x}^2 = \frac{1}{2m}(4\sum_{j=1}^n{\nabla_{\grave{x}^{2j-1}}\nabla_{\grave{x}^{2j}}}-\sum_{i=1}^d{\nabla^2_{x^i}}),\nonumber\\
&&\tilde{T}_0= \frac{1}{4}({\tilde x}\nabla_{\tilde x}+\nabla_{\tilde x} {\tilde x})=\frac{1}{2}(\sum_{i=1}^d{x^i\nabla_i}+\sum_{j=1}^{2n}{\grave{x}^j\nabla_{\grave{x}^j}})+\frac{\tilde{d}}{4},
\end{eqnarray}
such that the conformal algebra (\ref{algebra_sl2}) still holds.  Here $\tilde{d}\equiv d-2n$ is the super dimension.   
Our goal is to construct a holographic model dual to the quantum Calogero model with following Hamiltonian in superspace,
\begin{equation}
\tilde{\cal H}=\tilde{T}_- - \omega^2 \tilde{T}_+ + c \tilde{T}_0.
\end{equation}
To achieve that, we add to the superspace additional null-like directions $\tilde{t}=t e_t$ and $\tilde{\xi}=\xi e_{\xi}$, together with a holographic direction $\tilde{r}=re_r$, all with trivial anti-commuting components.  We also request in the extended Schr\"{o}dinger symmetry, the super coordinates scale as 
\begin{equation}
(\tilde{t},\tilde{\xi},\tilde{r},\tilde{x}^i) \to (\lambda^2\tilde{t},\tilde{\xi},\lambda^{-1}\tilde{r},\lambda\tilde{x}^i).
\end{equation}
The proposed a super geoemtry in $R^{d+3|2n}$ then simply reads,
\begin{equation}
d\tilde{s}^2 = \tilde{g}_{ij}d\tilde{x}^i d\tilde{x}^j,
\end{equation}
with exactly same metrics as in (\ref{metrics_one}) and (\ref{ansatz_one}) but with $\vec{x}$ replaced by $\tilde{x}$.  In summary, we have proposed a {\it super}geometry in $R^{d+3|2n}$ holographic dual to the one-particle super Calogero model in $d$-dimensional space.  It is straightforward to generalize this construction in $R^{Nd+3|2n}$ for $N$-particles.

\section{Discussion}
We have constructed a holographic gravity model dual to the non-relativistic (super) Calogero quantum model for single and multiple particles, as a generalization of the recent developments on AdS/NRCFT.  We have found that the addition of $dtd\vec{x}_i$ (i.e. function $B_i(r,\vec{x})$) to the metrics corresponds to turn on both harmonic traps as well as the dilatation term in the Calogero model.  A few remarks are in order.  First, we would like to comment on those free parameters inside our metric.  We have learnt that Hamiltonian's taking the forms,
\begin{equation}
{\cal H}_{\omega,c}=-T_- + \omega^2 T_+ + c T_-
\end{equation} 
are {\it equivalent} to each other up to a unitary transformation\cite{Meljanac:2005hm,Galajinsky:2006hq}.  For instant, the Hamltonian (\ref{hamiltonian_calogero}) with nontrivial $b_1$ can be mapped to that with $b_1=0$, denoting ${\cal H}'$, by a combination of transformations,
\begin{eqnarray}
&&{\cal H}_{\sqrt{2a_2+b_1^2},i2b_1}= S_1 S_2^{-1} {\cal H}'_{\sqrt{2a_2},0} (S_1 S_2^{-1})^{-1},\nonumber\\
&&S_i=e^{-\beta_iT_+}e^{-\alpha_iT_-},\qquad \alpha_1=\alpha_2=\frac{1}{\sqrt{2a_2}}, \qquad \beta_i=\sqrt{2a_2}-i\delta_{i1} b_1,
\end{eqnarray}
Therefore metric deformed by various choices of $a_i,b_i,\gamma$'s give rise to different definition of $\it energy$, which may be related by unitary transformation.

Secondly, we simply stated the desired super geoemtry to reproduce the super Calogero quantum model without knowing too much beyond the super metric.  We remark that the supersymmetric extension of the correspondence has been initiated  recently in \cite{Sakaguchi:2008rx}, and alternative supersymmetric extension of Calogero model has also been studied in \cite{Freedman:1990gd,Galajinsky:2007gm}.  It would be interesting to establish any connection between our proposal and their construction.

\begin{acknowledgments}
The authors are partially supported by 
the Taiwan's National Science Council and National Center 
for Theoretical Sciences under Grant No. NSC96-2811-M-002-018, 
NSC97-2119-M-002-001, and NSC96-2811-M-002-024.
\end{acknowledgments}


\bibliography{apssamp}

\end{document}